# Personal Information Leakage During Password Recovery of Internet Services

Mordechai Guri, Eyal Shemer, Dov Shirtz, Yuval Elovici
{gurim, eyalshem, dovshirtz, elovici}@post.bgu.ac.il
Ben-Gurion University of the Negev, Israel
Cyber-Security Labs

*Abstract* – **In this paper we examine the standard *password recovery* process of large Internet services such as Gmail, Facebook, and Twitter. Although most of these services try to maintain user privacy, with regard to registration information and other personal information provided by the user, we demonstrate that personal information can still be obtained by unauthorized individuals or attackers. This information includes the full (or partial) email address, phone number, friends list, address, etc. We examine different scenarios and demonstrate how the details revealed in the password recovery process can be used to deduct more focused information about users.**

## I. INTRODUCTION

In the past decade, Internet services have become an integral part of life. With electronic mail services, shopping websites and social networks, users enjoy an unprecedented ability to make transactions online. In order to facilitate this, users are requested to provide specific information and personal details, which may vary depending on the service. For example, users voluntarily share their credit card details, full name, phone number, and home address. Users also engage in private conversations in social networks and upload personal photographs to share them with others on the network. In order to access a particular service, users go through a process of identification and authentication in which users identify themselves by providing their username, and authenticate this with a password. Services are prepared, however, for a common scenario, in which users forget their passwords. In an effort to allow users to recover their passwords, services use different techniques and mechanisms to allow secure password recovery. While efforts to ensure that the true owner of an account is able to log in are important, most types of password recovery processes are vulnerable to information leaks. Information leaks can occur whether password recovery is handled by email, phone, or other alternatives, and when they do, elements of the user's private information can potentially be revealed to other interested parties. In this paper we examine a few major Internet services and their password recovery mechanisms. We compare them and present realistic scenarios which demonstrate the potential personal information leaks which could accompany these situations.

## II. PERSONAL INFORMATION LEAKGE

To demonstrate the potential information leakage during the password recovery process and its possible implications, we consider the following fictitious user, John Smith and present, step by step, an attack scenario involving this user. The attacker doesn't have any prior knowledge about him, other than his name and his email address, *johnspersonalmail@gmail.com*.

**1. Alternative Email:** The attacker starts with Facebook. Using Facebook's password recovery mechanism described above, the attacker obtains parts of Smith's alternative email address, in the format of *******@f*****.edu. This information has numerous derivatives. For example, the "edu" in the address suggests that Smith has an academic background.

**2. Friends' Identities:** Since Facebook also implements a unique friend verification process, the attacker then attempts to use it for John Smith's profile. During the process, the attacker obtains three groups of friends from Smith's life. In an effort to make this process harder for attackers, the three groups of friends provided by Facebook often come from different aspects of the user's life. The attacker notices (by visiting the friends' Facebook profiles) that Smith's first group of friends includes many Americans with an academic background, many of whom are currently undergraduates at universities.

**3. Age and Education**: Based on the information obtained, the attacker reasonably assumes that Smith is between the ages of 18 and 22. By scrutinizing the user's friends, the attacker observes that the most dominant academic institution is Furman University, which fits perfectly with Smith's email address. This leads to the conclusion that Smith is currently a student at Furman University and hence, resides in Greenville, South Carolina. The second group of friends includes many people from Raleigh, North Carolina; therefore, the attacker suspects that Smith's hometown is Raleigh. The third group of friends includes many people whose profile pictures and public profiles suggest some degree of animal rights activism; the attacker thus concludes that Smith is involved with, or supportive of, animal rights activism.

**4. Phone number:** In the next step, the attacker considers the only other piece of information initially known about John: his email address. In Gmail's password recovery process, the attacker will then find out that John's phone number is in the format of *********22; however, the attacker is now aware that John is originally from Raleigh, North Carolina. The attacker knows that Raleigh's area code is 919, which means John's number is 919-***-**22. There are $10^{10}$ potential phone numbers for a ten-digit number.

**5. Personal Identity:** The attacker has now narrowed down the potential numbers to $10^5 = 100,000$ numbers, which is exponentially lower. This information makes it much easier to determine Smith's complete phone number using search in public phone directories.

In virtually no time, our attacker has formed a partial profile of John Smith. He attends Furman University. He was born in Raleigh, North Carolina and attends college at Furman University. His age is between 18 and 22. He is involved with animal rights activism, and his phone number is *919-***-**22*.

### III. SERVICES

In order to evaluate potential information leakage on social networks, we mapped various Internet services that hold valuable user information. After mapping these services, we examined the services' password recovery mechanisms (also referred as "password reset"). While some of the services preserve user privacy, others allow information leakage. The mapping process for each service includes four main steps: (1) inserting username/password, (2) selecting the password recovery option, (3) following the password recovery process, and (4) if the process includes alternatives, following the alternations. During steps 3 and 4 we record every instance of personal information leakage.

#### A. Facebook

Facebook has a two-phase recovery mechanism. The first and easier phase allows the user to perform password recovery using alternative, pre-defined user accounts. These accounts include the user's (1) email address, (2) Google account, or (3) mobile phone number for password reset. When the user is initially identified, their full name is usually displayed under their image. Some additional information may be displayed such as the user's academic institution, gender, or age - depending on the user's privacy preferences. The ability to reset the password by electronic mail also reveals parts of the user's mail address. Depending on the domain, some of the username is provided (for example, first and last letter). Parts of the domain may also be displayed, and in some cases, the whole domain will be visible. In order to properly identify the user's smartphone device, the last three digits of the phone number are presented. In this way, Facebook reveals valuable pieces of information about the user.

Facebook provides another option to authenticate a user in the case of a lost password. This technique relies on the fact that the user's friends can provide another form of authentication. In practice, this happens in two steps. First, the user is presented with three groups of his/her friends; each group contains tens of friends. The user has to choose one friend from each group. Then, verification codes are sent to each of the chosen friends, with a warning discouraging the friends to provide the code before talking to the user first. Effectively, the responsibility to authenticate the user is given to his/her friends. If the user successfully obtains these three codes, access to the account is granted.

This form of authentication is unique to social networks, however it also leaks potentially important information about the user, by granting access to the user's friends. A potential attacker can easily obtain this information and abuse it. If the attacker is an acquaintance of the user, it may be possible for them to contact a friend from each group and obtain the verification codes. It should be noted, however, that Facebook generates these three groups in a way that takes into account a certain modularity of the user's friends' social networks. For instance, the user may be presented with groups of friends from high school, university, and the workplace. While it is unlikely for an attacker to know a member of each of the three groups of the user's friends, it is possible. It is not safe to assume that the user's friends will behave ethically, or even care about the user's well-being; hence, a situation where the verification codes are given to an attacker is reasonable.

#### B. Gmail

Gmail also provides a simple password recovery mechanism. Given the assumption that the only information available is a proper identification (i.e., a username), the user is requested to enter one of his/her previous known passwords. If the user skips this step, Gmail suggests the first form of verification, by phone. If no access by phone is possible, the next form of verification suggested is via an alternative email account. Given a situation in which the user has no previous knowledge of an alternative mail address, previous passwords, or a phone number, Gmail does not offer any further means of password recovery. During these steps, Gmail reveals some private information about the user. In the attempt to identify the user's phone number, the last two to three digits *(*depending on the user's phone number) of the stored phone number are revealed. In the case of email verification, some parts of the user's alternative email address are revealed. This includes the first and last letter of the username and parts of the domain name. These two types of data are the only private information leaked by Gmail. Provided that some sort of account identification must be made, this information leak is minimal, and GMail's password recovery mechanism is relatively secure.

#### C. PayPal

The password recovery mechanism implemented by PayPal relies on access to the email address associated with the account. Provided that no such access exists, no information is leaked. Upon providing the email address, a verification mail is sent. After this email is opened, a redirection link to PayPal will appear. The user will then be asked to choose and confirm a new password, along with selecting and answering two security questions. Following this quick procedure, full access to the account is granted. In addition, PayPal allows password recovery based on phone confirmation. Similar to other services, PayPal provides a hint by revealing the three last digits of the phone number. While PayPal does not leak any other information, access to an associated email account or phone number can allow someone else to take over an account.

*D. Other Services*

A Microsoft account allows access to different Microsoft services, such as Office and Skype. Some of the private information that can be mined includes chat conversations with contacts and various Microsoft licenses. Microsoft also implements a simple password recovery process; when a user clicks on 'Forgot my Password,' he/she will be asked to provide a registered email address or a phone number for email/SMS verification. During this process, no personal email information is revealed; however, password recovery via SMS leaks the last two digits of the user's phone number. Other services such as Yahoo! and Twitter reveal some personal information as well. For Yahoo!, the password recovery process divulges the last two digits of the user's phone number. For Twitter, in the case in which the user successfully inserts a username and/or phone number, parts of the email address will be revealed: the two letters of the username, along with the first letter of the email's domain. Table 1 provides a summary of the results, listing the services evaluated and the information leaked.

| SERVICE | INFORMATION |
|---|---|
| **Facebook** | Last two digits of phone number |
| | Parts of associated email address |
| | Subsets of user's friends (randomly generated) |
| **Gmail** | Last two to three digits of phone number |
| **PayPal** | Last three digits of phone number |
| | Parts of associated email address |
| **Twitter** | parts of email address |
| | The first two letters of email username |
| | The first letter of the email domain name |
| **Yahoo!** | Last two digits of phone number |
| **Microsoft** | Last two digits of phone number |

**Table 1: Summary of the results.**

IV. ADDITIONAL ATTACKS

The email account and mobile phone number associated with the services play an important role in the password recovery process. By obtaining access to an email account or mobile phone, an attacker can actually gain access to other services as well. In the follow subsections we describe two possible attack scenarios which involve access to email account in user mobile phone.

*A. SMS Recovery Attack*

Given the reasonable assumption that no access to the mobile phone is granted to the attacker, we present a scenario where strategic information within the phone can be obtained. For example, assume a user has an Android OS smartphone. The attacker creates a seemingly harmless application (e.g., an application that measures and reports the distance a user walks daily, provides notifications about walking and nearby friends, etc.). Such an application may require certain permissions that would not arouse suspicion given its functionality: reading and receiving SMSs, Internet/network access, etc. Once the user downloads the malicious app, the details are recorded by the attacker. Then, at a random time, the malicious app secretly performs the password recovery process by SMS. Services that allow phone password recovery (such as Gmail) are vulnerable to this type of attack as follows. When an SMS is received by the user's phone with the given verification code, the message is caught by the application which sends the message to a remote server. At this point, the attacker can use the verification code and complete the password recovery, resulting in full access to the account.

*B. Email Recovery Attack*

This attack uses a similar technique, but instead of exploiting the SMS, it exploits the email recovery option. In this case, the malicious app needs specific permissions to enable it to read received mail. As in the previous example, the attacker will go through the email verification process and silently catch an incoming mail and send it to a remote server where the mail can be analyzed; then the attacker can obtain the verification code, perform a password reset, and obtain complete access to the given service.

V. RELATED WORKS

The intention of this brief survey is to present some widely used password recovery mechanisms and point out their vulnerabilities with regard to information leakage. The most common methods for password reset and recovery rely upon the use of predefined email addresses, security questions, and telephone numbers which are used to send verification SMSs or email messages or receive a call from the service's helpdesk. Another recent method uses the human relationship [10]. A simple authentication method for the users' password reset case is suggested by [11]. Authentication of users is done with the use of security token generated and based on sending security tokens via email and a link that enables the user to perform its password reset activity. However, [13] point out that email based attacks have been largely unsuccessful, and they suggest investing more thorough methods of user authentication by phone. The mechanism of using a serious question for user verification in the password recovery process has been presented by Ellison et al. [3]. However, [4] showed that the suggested mechanism is quite weak and is not advisable in all cases. An analysis of the security and reliability of secret questions or "life-questions" that are used to authenticate the user in the password reset process of the largest email providers (AOL, Google, Microsoft, and Yahoo!) is presented in [12]. The research shows that users tend to forget their answers to these questions. As a result of this research, Yahoo! changed its questions in February 2009. Kumar states that a weak password recovery/change mechanism based on "life question" is quite weak; perhaps because the security questions are easy or common, the answers are likely to be guessed by an unauthorized person. Another weakness pointed out by Kumar is the possibility of sending a new password to an email that was planted by the attacker–not the legitimate user [6]. Grossman pointed that password recovery mechanisms that are not strong enough may jeopardize the website, and therefore expose a firm and its management to legal procedures and claims. His suggestion is to use an email address for the user verification process [5]. Irani et al. show that a password recovery attack on social networks may reveal 38%-90% of the users' private data [2]. Having some knowledge about the target is helpful, since many services still use a method that involves answering some predefined personal questions for verification for password recovery, such knowledge makes it easy to reveal private information regarding the attacked personnel [1]. Password recovery processes using a "secret question" has been discussed by many researchers. [14], [15], [16], and many have claimed that the password recovery processes that use a "secret question," are likely insecure. The reason for this is because the answer to the question may be public knowledge. For example, if a hacker knows the victim's name and some historical information about him/her, the attacker may know the answer to the "what is the name of your elementary school" question [16].

## VI. Conclusion

This work examines password recovery and password reset processes of several major Internet services. We demonstrate how some pieces of private information might be exposed during these processes. Finally, we propose scenarios where such information could be collected (in full or partially) and cross-checked to target identification about a specific user. As already discussed in [6] and other research, password recovery mechanisms should be carefully designed and implemented. Otherwise they can be misused, enabling an attacker to access users' private information.